\documentclass[trackchanges,twocolumn]{aastex701}
\usepackage{graphicx} 
\usepackage{longtable}
\usepackage{appendix}
\usepackage{url}
\usepackage{booktabs}
\usepackage{multirow}
\usepackage{amsmath,amssymb}
\usepackage[table]{xcolor}
\usepackage{hyperref}
\usepackage{amsfonts}
\usepackage{wrapfig}
\usepackage[normalem]{ulem}

\begin{document}

\title{Using gravitational lensing to probe for bright quintessential galaxies in the Epoch of Reionization}

\author[orcid=0000-0002-0975-623X]{Joshua Roberson}
\affiliation{Department of Physics, University of Cincinnati, PO Box 210011, Cincinnati, OH 45221}
\email{robersju@mail.uc.edu}

\author[orcid=0000-0003-1074-4807]{Matthew B. Bayliss}
\affiliation{Department of Physics, University of Cincinnati, PO Box 210011, Cincinnati, OH 45221}
\email{baylismb@ucmail.uc.edu}

\author[orcid=0000-0003-1370-5010]{M.D. Gladders}
\affiliation{Department of Astronomy and Astrophysics, University of Chicago, 5640 South Ellis Avenue, Chicago, IL 60637}
\email{gladders@astro.uchicago.edu}

\author[orcid=0000-0002-3475-7648]{Gourav Khullar}
\affiliation{Department of Astronomy, University of Washington, Box 351580, Seattle, WA 98195}
\email{gkhullar@uw.edu}

\author[orcid=0000-0002-7559-0864]{Keren Sharon}
\affiliation{Department of Astronomy, University of Michigan, 1085 S. University, Ann Arbor, MI 48109}
\email{kerens@umich.edu}

\author[orcid=0000-0001-6505-0293]{Keunho Kim}
\affiliation{Division of Physics, Mathematics, and Astronomy, California Institute of Technology, 1200 E California Boulevard, Pasadena, CA 91125}
\email{keunho11@caltech.edu}

\author[orcid=0000-0001-7410-7669]{Dan Coe}
\affiliation{Space Telescope Science Institute, 3700 San Martin Drive, Baltimore, MD 21218}
\email{dcoe@stsci.edu}

\author[orcid=0000-0001-7665-5079]{Lindsey Bleem}
\affiliation{High Energy Physics Division, Argonne National Laboratory, 9700 S. Cass Avenue, Lemont, IL 60439}
\email{lbleem@anl.gov}

\author[orcid=0000-0003-3521-3631]{Taweewat Somboonpanyakul}
\affiliation{Department of Physics, Faculty of Science, Chulalongkorn University, 254 Phayathai Road, Pathumwan, Bangkok 10330, Thailand}
\email{Taweewat.S@chula.ac.th}

\begin{abstract}
\noindent Understanding the properties of the first generation of galaxies is an ongoing challenge in observational astrophysics. While advances in deep field observation have led to the identification of large numbers of galaxies within the Epoch of Reionization, there are very few observed galaxies at this range that are sufficiently bright for high signal-to-noise spectroscopy. To this end, we analyse HST and ground-based photometry of five candidate strongly lensed galaxies, all projected behind the cores of massive clusters and with similarly red optical-NIR colors. All are characterized by a drop-off in their spectra between the near-infrared and optical wavelengths, corresponding to a Lyman-break that sets a lower bound on their redshifts. Using the open-source SED modeling software Prospector, we characterize two of these galaxies as high-$z$ ($z \sim$ 6.5-7) while the other three are low-$z$ ($z \sim$ 2) despite all five having similar apparent magnitudes at the observed wavelengths. We demonstrate that for the brightest dropout candidates we can distinguish high-$z$ galaxies from red or dusty low-$z$ galaxies using limited photometric data. The bright sources enable deep constraints on the dropout color which, in combination with flat continua measured in redder bands, require high-$z$ solutions when searching the parameter space. At the time of writing this work significantly increases the number of m$_{AB}$ $<$ 24 galaxies at or above a redshift of 6, and provides a path forward for future analysis on the early era of galaxy formation.
\end{abstract}


\section{Introduction}\label{section:intro}

\noindent The Epoch of Reionization (EoR) stands as the transitional phase that gave rise to the current inhomogeneous state of our Universe, wherein the quasar and stellar populations that would become the first galaxies reionized overdense regions of hydrogen gas \citep{Loeb01}. Over the past two decades, analysis of various cosmic microwave background anisotropies, Lyman-alpha emitters, and active galactic nuclei has constrained the EoR to redshift $6 < z < 9$ \citep{Planck15, Planck16,Mason19,Jin23,Spina24,Raghunathan24}. Over the course of this era, the cosmic fraction of neutral hydrogen gas drops to nearly 0; however, the \textit{cause} of this change is directly tied to the properties of the very first stars and galaxies - properties that are still not well understood. To this end, modeling and categorizing reionization-era galaxies is an active field, pushed forward by an increasingly large catalogue of high-redshift galaxies characterized both photometrically and spectroscopically \citep{Bowler12,Bradley14,Bouwens15,Finkelstein15,Ishigaki18,Salmon20,Endlsey22}. The launch of the \textit{James Webb Space Telescope} (JWST) has already begun to extend this space even further, with large-scale surveys doing deep field observations at redshift ranges previously unattainable \citep{eisenstein23,Noirot23,finkelstein23,withers23,tripodi24,price24}.
\newline

\noindent However, even as the high-z space expands rapidly, there remain limitations on the amount of meaningful analysis that can be done to characterize these galaxies within the context of the EoR. One of the prevailing questions surrounding reionization is that of the primary driving source: was it a small number of high-mass galaxies, a larger number of low-mass galaxies, or active galactic nuclei (AGN)? Although literature in the last decade has been trending towards the second interpretation \citep[e.g.;][]{Atek15,Livermore17,Atek24EoR,Simmonds24}, new in-depth observations of galaxies, AGN, black holes, and other ionizing sources with \textit{JWST} have shown that we do not have a complete picture of the ionization process \citep[e.g.;][]{Fujimoto23,dayal24,Greene24,Madau24}. While this question can be answered in part by characterizing the luminosities of EoR galaxies, a problem of statistics immediately arises: the high-mass galaxies that are the brightest in the optical and infrared wavelengths and thus more suitable targets for a full spectroscopic analysis are still much rarer compared to their dimmer counterparts, and the self-same characteristics that make them easy to observe means that they are non-representative of the lower-mass galaxies thought to be the main drivers of reionization. Conversely, while the dim lower-mass drivers are much more numerous, obtaining spectroscopically meaningful high signal-to-noise data for them is significantly harder and more expensive; while recent \textit{JWST} surveys have begun to explore this space in earnest \citep[e.g.;][]{Qiao24,Mascia24}, measuring the properties of extremely faint galaxies still remains a challenge even with \textit{JWST} observations. 
\newline

\noindent Measuring the detailed properties of faint, high-redshift galaxies requires studying the most highly magnified, strongly-lensed sources in the distant universe. Strong gravitational lensing transforms the small and faint galaxies more typical of the high-redshift space into highly magnified images in the sky that are observationally bright enough to do high signal-to-noise analysis where it would otherwise be impractical. In particular, the brightest, highly magnified galaxies are extremely rare but hold significant scientific value, as they represent intrinsically faint sources that can be studied in much greater detail (e.g., isolating nebular lines from high-resolution spectroscopy) than would otherwise be possible. It is clear from the basic shape of the galaxy luminosity function at high redshift ($z \gtrsim 6$) that faint galaxies are far more numerous than their bright counterparts \citep{Finkelstein15, bouwens21}, and as such measuring the detailed properties of faint high-redshift galaxies is crucial for understanding how galaxies were assembled and how they contributed to re-ionizing the universe. \cite{Atek24EoR} is one early \textit{JWST}-era example of in-depth spectroscopic analysis of these highly magnified galaxies and shows that they are strong sources of ionizing photons in agreement with earlier literature; however, the physical properties of the interstellar medium that would actually produce said photons are still unclear, suggesting a need for more examples.  
\newline

\noindent Finding this class of highly-lensed galaxies requires space-based imaging surveys of known lensing sources, typically galaxy clusters. These surveys, while observationally expensive, have been successful at capturing and characterizing galaxies from anywhere as close as $ z \sim 1$ to as far as the so-called `Cosmic Dawn' period of the first galaxies. Previous \textit{Hubble} surveys of known strong lenses and their associated lensing cross-sections such as the Cluster Lensing and Supernova Survey with \textit{Hubble} (CLASH; \citet{Postman12}, Hubble Frontier Fields (HFF; \citet{Lotz17} and the Reionization Lensing Cluster Survey (RELICS; \citet{Coe19paper} have produced extensive catalogues of lensed galaxies (e.g. \citet{bouwens22}, \citet{mestric22}, and \citet{bergamini23} and in-depth analyses of particularly noteworthy targets (e.g. \citet{Coe13}, \citet{Salmon18}, or \citet{Calura21}. 
\newline

\noindent However, while this space-based approach excels at finding faint high-redshift galaxies, it relies on targeting specific clusters and thus can only cover particular narrow ranges of the sky. In contrast, ground-based surveys such as the SDSS Giant and Bright Arc Surveys (SGAS, \citet{Sharon20}; SBAS, \citet{Kubo09} and CASSOWARY \citep{cass14} which search wide areas of the sky find the brightest strong lensing systems as a result of their wider field of view, with the latest generation of wide-field ground-based imaging surveys producing even more samples of bright strongly lensed systems (DECaLS, \citet{Huang20}; DES, \citet{Jacobs19}; SuGOHI-c, \citet{Jaelani20}. While this ground-based approach is more shallow and thus mostly identifies lensed galaxies $1 < z < 4$ \citep{Bayliss_2011a,Bayliss_2011}, the increased area coverage and thus higher volume of lensing systems viewed across more lines of sight means that especially rare bright high-redshift galaxies are found as targets. 
\newline
    
\noindent It follows that a viable search approach lies within the intersection of these two approaches: focused space-based observations to find the highest redshift sources, cross-referenced against wide ground-based surveys to find the brightest sources. Here, we present the discovery and initial parameterizations of five bright, candidate strongly lensed high-redshift galaxies identified with this methodology, two of which we photometrically constrained to be in the $ z \sim $ 7 range. This paper is structured as follows. Section 2 details the observation and search process that led to the discovery of these galaxies. Section 3 outlines the follow-up observations that we obtained for the sample of candidate high-redshift galaxies that resulted from our search. Section 4 describes the calibration and photometric spectral energy distribution modeling that allows us to characterize these galaxies, and places the two high-redshift galaxies in the wider context of the $z >$ 5 space. Section 5 serves as a summary of the paper and details the future work intended for these galaxies and the methodologies described in this paper. Magnitudes are reported calibrated to the AB system. The underlying cosmology model for all simulations assumes a standard flat, cold dark matter universe with cosmological constants based on WMAP9 observations: $\Omega_{\Lambda} = 0.721 \pm 0.025, H_0 = 70.0 \pm 2.2$ \citep{Hinshaw13}. All parameters reported are the 50th percentile confidence interval values with uncertainties reported as the 16th and 84th percentile values, unless noted otherwise. 

\section{Mining for Bright Dropout Candidates}

\noindent We performed a photometric search for bright ($m_{\rm AB} \lesssim 25$) candidate high-redshift dropout galaxies (a galaxy wherein the luminosity sharply falls off below a specific wavelength) in fields containing massive galaxy clusters that contained both optical and near-infrared (NIR) imaging from \textit{HST}, but were not searched for in previous programs such as CLASH \citep{Postman12}, the HFF \citep{Lotz17}, or RELICS \citep{Coe19paper} that also looked for bright lensed galaxies. The focus on searching massive cluster fields was motivated by the possibility of finding very rare, highly magnified galaxies at high redshift. In practice, this search involved careful visual inspection of the multi-band \textit{HST} imaging for each field to identify very red, bright sources seen near the cores of the foreground massive clusters. 
\newline

\noindent We conducted this search in the fields of 37 galaxy clusters from the Sloan Giant Arcs Survey (SGAS; \cite{Bayliss11,Sharon20}, imaged in the large program HST-GO-13003 (PI: Gladders), and 137 galaxy clusters from two South Pole Telescope (SPT) surveys: the SPT Sunyaev Zel`dovich 2500 deg$^{2}$ survey (SPT-SZ; \cite{Bleem15} and the SPT Extended Cluster Survey (SPT-ECS; \cite{Bleem20}, imaged across one Snapshot program over two HST cycles (Cycle 25 HST-GO-15307 and Cycle 26 HST-GO-16017; PI: Gladders). These cluster samples substantially expand the number of lines of sight toward massive clusters that have been searched for brightly lensed sources -- together, CLASH, HFF, and RELICS include fewer than 70 massive cluster lines of sight, while the SGAS and SPT cluster samples searched here include 174 new, independent lines of sight. All of the \textit{HST} data used in this paper can be found in MAST: \dataset[10.17909/qyf8-t691]{http://dx.doi.org/10.17909/qyf8-t691}

\subsection{SGAS-HST Archival Data}

\noindent The SGAS galaxy clusters are identified as strong lenses based on the presence of at least one giant arc in shallow ground-based imaging data. The GO-13003 data is a subset of 37 galaxy clusters with two WFC3/IR bands for each cluster, one of which is always F160W and the second of which is either F105W or F110W. Each cluster was also observed in two WFC3/UVIS filters that varied from cluster to cluster; the bluer UVIS filter was always either F475W or F390W and the redder filter one of F814W, F775W or F606W. The exact filter choices for each cluster field were motivated by optimizing the placement of spectral breaks in known, bright giant arcs that appear in each field. For more details about the SGAS-HST cluster sample and the complete HST-GO-13003 dataset and its reduction we refer the reader to \citet{Sharon20}.

\subsection{SPT Snapshot Archival Data}\label{hstsnap}

\noindent The SPT clusters were observed as a part of two HST Snapshot programs (Cycle 25 HST-GO-15307 and Cycle 26 HST-GO-16017; PI: Gladders) that observed 137 galaxy clusters from two South Pole Telescope (SPT) surveys: SPT-SZ \citep{Bleem15} and SPT-ECS \citep{Bleem20}. The observed clusters were drawn randomly based on available windows in the HST schedule, from a parent sample selected such that clusters had an equal co-moving number density at $ 0.2 < z < 1$. All the clusters at $ z \ge 1 $ were included, and a lower redshift limit was applied at $ z = 0.2 $ due to the small volume this redshift bin samples and the extensive data already available for low-redshift clusters. A redshift-dependent lower limit mass threshold was selected, anchored at $ M_{500\mathrm{c}} = 4\times10^{14} M_{\odot} h^{-1} $ at the median catalogue redshift. This threshold selects lower-mass clusters at higher redshifts in order to foster evolutionary studies: the less-massive high redshift systems evolve with cosmic time to become the high-mass systems at low redshifts. The final parent sample included 293 clusters that met these criteria and had no pre-existing HST archival data; 137 of the 293 were observed in the two HST Snapshot programs. 
\newline

\noindent The Snapshot imaging of SPT clusters was obtained with the WFC3-IR camera through the F110W filter, and with the WFC3-UVIS camera through the F200LP filter, an ultra-broad filter that essentially samples the full UVIS detector quantum efficiency from $\sim$2000 \AA\ to $\sim$1 $\mu$m. Each cluster was observed in these two bands in a single shortened SNAP orbit. These two filters were selected in order to optimize the throughput during a partial HST orbit, as they provide the broadest UVIS or IR wavelength coverage. The observations were configured to have one of three total exposure lengths (short, medium, or long) depending on cluster redshift, where higher redshift clusters were observed with a longer exposure time. The imaging in each filter was divided into three individual exposures using the standard subpixel 3-point line dither pattern. 

\subsection{SPT Snapshot Data Reduction}

\noindent Prior to running any pipeline reductions of the Snapshot data we first applied two corrections to the individual exposure images downloaded from MAST. The first correction addresses a visible mismatch in sky levels in the F200LP data between the two WFC3-UVIS chips. We rectify this issue by subtracting the background sky on each chip, with the sky estimated by fitting an iterated Gaussian model, sigma-clipped on the high end, to the pixel values. This correction is purely cosmetic with no effect on photometry, and is done to provide the best resulting data for constructing two-band pseudo-color images for further visual examination.
\newline

\noindent The second correction focused specifically on `IR blobs' in the IR/F110W data, which show up in the same locations across all images and in many cases overlapped with target objects such as cluster member galaxies and lensed arcs observed in the data. Under normal circumstances, the pixels affected by these features are flagged and not used in the final data product; however, due to the truncated SNAP orbit length the data are minimal in this program. It was thus deemed relevant to recover the areas affected by these IR blobs by careful flat-fielding. To do so we acquired several hundred additional F110W images of similar integration time from the archive (GO-15163), which we then combined with most of the F110W data described here using the IRAF  \textit{imcombine} task. We then used GALFIT \citep{galfit1,galfit3} to build a model of each identified IR blob across the stacked image from a combination of Sersic profiles. The composite full image of all the fitted IR blob models was then used to flat-field these features in the F110W data.
\newline

\noindent Final imaging data products for both F200LP and F110W data were generated following the standard procedures established by STScI, employing the Python3 package \textit{Drizzlepac} \citep{drizzle}. For astrometric matching of images, sub-exposures taken with the same filter were aligned to the same reference grid, using \textit{tweakreg} operating on tuned catalogues extracted using \textit{SExtractor} \citep{SEx}. The exposures were subsequently combined using the \textit{astrodrizzle} function of \textit{Drizzlepac} using a Gaussian kernel, with nominal cosmic ray rejection, a final pixfrac of 1.0 and 0.8 (for the F110W and F200LP, respectively) and pixel scale of 0.03" pixel$^{-1}$. 

\subsection{Dropout Selection}

\noindent Due to the crowded fields in the cores of massive clusters and the non-standard morphologies that can manifest for strongly lensed sources, push-button catalogue photometry can be subject to systematic biases. We had a modest (N$=$174) number of fields to examine, and our motivation was to find the rarest, most exceptional sources without concern for defining a sample in terms of completeness. We therefore performed our search by first visually inspecting the archival data described above for very red, bright sources around the cluster cores and then manually measuring photometry for the handful of scientifically significant red sources that emerged from visual inspection. 
\newline

\begin{table*}
  \centering
  \caption{Dropout Color Criteria}
  \begin{tabular}{lccccc} \toprule
        Cluster Field&  Blue Filter&  $m_{AB,Blue}$
         &  Red Filter&  $m_{AB,Red}$& Color\\\midrule
         SDSS~J0952$+$3434&  F105W&  25.43(13)&  F160W&           23.45(3)& 1.98\\
         SDSS~J0952$+$3434&  F125W&  25.06(3)&  F160W&  23.94(2)& 1.12\\
         SDSS~J0952$+$3434&  F105W&  26.43(42)&  F160W&  24.23(3)& 2.20\\
         SPT-CL~J0216-2609&  F200LP&  $>$28.69&  F110W&  23.87(7)& $>$4.82\\ 
         SPT-CL~J0304-5404& F200LP& $>$28.78& F110W& 23.87(10)&$>$4.91\\ \bottomrule
  \end{tabular}
\\[5pt]Brightness and color selection criteria for the five dropouts. Some sources are multiply imaged in the source plane; in those cases, the coordinates for the brightest image are reported.
  \label{tab:select}
\end{table*}

\noindent We then applied a set of criteria for both brightness and color. The specific brightness criterion for candidate selection was having at least one NIR band with an AB magnitude $m_{\rm AB} \leq 25$, and the color criterion was an AB $m_{\rm blue} - m_{\rm red}$ color $\geq1$ in adjacent filters. The search resulted in five bright, red candidates located in three cluster fields: SDSS J0952+3434, SPT-CL~J0216$-$2609 and SPT-CL~J0304$-$5404. All five candidates were measured to have total AB magnitudes in either F110W (SPT Snapshot data) or F160W (SGAS-HST data) of $\sim$ 24. Somewhat surprisingly, three candidates appeared near the core of a single cluster, SDSS J0952+3434, though with different colors, suggesting that they are separate sources rather than three images of a single source. From here out we refer to the dropout candidates in SDSS J0952$+$3434 as SDSS0952-DO-1, SDSS0952-DO-2, and SDSS0952-DO-3. The other two dropout candidates were identified near two separate galaxy clusters from the SPT sample: SPT-CLJ0216$-$2609 and SPT-CL~J0304-5404. We refer to these candidates as SPT0216-DO and SPT0304-DO. The basic properties used to select all five dropout candidates are summarized in Table~\ref{tab:select}.
\newline

\noindent We emphasize two specific and readily apparent galaxy properties as the selection criteria for these five galaxies from the wider galaxy sample. The first is a combination of bright apparent flux and close proximity to a massive galaxy cluster core, giving them a high chance to be gravitationally lensed and highly magnified. The second is a sharp drop-off, or spectral break, in adjacent band-passes, which is a relatively simple observational signature that is well-known as a method of identifying Lyman-break galaxies \citep{Steidel96}. This method of selection has been used in multiple surveys to identify high-redshift ($z \gtrsim 6$) galaxies with a high degree of confidence by searching for spectral breaks redward of $\sim$8500 \AA. Generally, interloper sources such as low-redshift galaxies with prominent dust attenuation, galaxies with very old stellar populations, or even local M dwarf stars can pose problems as they can masquerade as candidate high-redshift dropouts \citep{Salmon20}. For this set of galaxies, all of the sources we identify here are clearly extended and therefore cannot be foreground stars or brown dwarfs; however, it was necessary to determine whether or not they were low-$z$ interlopers. 
\newline

\section{Follow-Up Observations}

\noindent In order to rigorously test all five dropout candidates and distinguish them between high-confidence high-redshift objects and low-redshift interlopers, we collected additional archival and follow-up imaging and photometry across various band-passes not used in the discovery stage. 

\subsection{Spitzer/IRAC Imaging}

\noindent Two of the fields containing dropout candidates have archival imaging from the IRAC instrument on the Spitzer Space Telescope. SDSSJ0952$+$3434 has IRAC Channel 1 (Ch1; $\lambda_{\rm Eff} \sim 3.6\mu$m) and Channel 2 (Ch2; $\lambda_{\rm Eff} \sim 4.5\mu$m) data taken in Program \#70154 (PI: Gladders). The individual frame integration times were 30 s; the total per-pixel integration times near the center of the mosaic were 7.9 minutes in IRAC channel 1 (3.6 $\mu m$) and 8.9 minutes in IRAC channel 2 (4.5 $\mu \text{m}$). We downloaded and reduced these data with AOR 40796160 using the same method described in Section 3.2 of \citet{Florian_2021}. SPT-CL~J0304$-$5404 has IRAC Ch1 and Ch2 data taken in Program \#80012 (PI: Brodwin); we use the same reduced version of these data presented in \citet{Bleem15}, and we refer the reader there for more details on the reduction.

\subsection{New HST Imaging for SDSSJ0952$+$3434}

\noindent We obtained new imaging data for the field of galaxy cluster SDSSJ0952$+$3434 to better constrain the photometric redshifts of bright candidate high-z dropout galaxies identified near the cluster core. New HST imaging was taken in program HST-GO-14896 (PI: Bayliss) with WFC3/IR in the F098M, F125W, and F140W broadband filters. Imaging in all three filters was performed in a 4-point dither pattern with subpixel dithers to enable robust reconstruction of the point spread function (PSF). The total integration times were 1862s in F098M, 1712s in F125W, and 1812s in F140W. These data were combined with the extant imaging in F606W, F105W and F160W to densely sample the spectral region around the apparent photometric break and the shape of the spectral slope between 0.6-1.6 $\mu$m. The data were reduced using the \texttt{drizzlepac} software package using the same approach as the HST data from program GO-13003 described in \citet{Sharon20}.

\subsection{Magellan/LDSS3c z-band Imaging}

\noindent We obtained imaging in the $z-$band filter with LDSS3c the on Magellan-II (Clay) telescope on January 16, 2021 of SPT0216-DO and SPT0304-DO between UT $\sim$01:00-02:00 and $\sim$03:00-03:50, respectively. Observations for SPT0216-DO consisted of 13 dithered 240 s exposures, while for SPT0216-DO we took 11 dithered 240 s exposures. We reduced these data using standard \texttt{IRAF} scripts from the \texttt{noao.imred.ccdred} package to bias-correct, flat-field, and align the individual science exposures. The final stacked $z-$band image for SPT0216-DO has a total integration time of 3120 s and a median seeing of 0.8". The final stacked $z-$band image for SPT0304-DO has a total integration time of 2640 s and a median seeing of 1.1". 
\newline

\noindent We measure the photometric calibration for the images of both SPT0216-DO and SPT0304-DO by comparing the LDSS3c $z-$band instrumental magnitudes against published $z-$band magnitudes from the PISCO instrument on Magellan \citep{Stalder14} and from the Dark Energy Survey \citep[DES;][]{abbott2021}. In both fields we use stars with AB magnitudes between $\sim$20-22.5 to compute the zero-points in the LDSS3c data. The red end sensitivity of the $z-$band filter transmission curves differ substantially across LDSS3c, PISCO, and DES/DECam, such that we expect a non-negligible color term in our photometric calibration step. We use the two different calibration references (PISCO and DECam) to make two independent estimates of the calibration, including the color terms. The final uncertainty in the $z-$band zeropoints, including uncertainty estimated in the color terms, is 0.08 magnitudes for each field.

\subsection{Magellan/FOURSTAR H-band Imaging}

\noindent We imaged SPT0216-DO and SPT0304-DO in the H and K$_{\rm s}$ bands with FourStar on the Magellan-I (Baade) telescope on the nights of October 16 and 17, 2019. These observations were performed using many random dither positions, spending $<90$s at each individual dither position, and tailoring individual exposure times to ensure that the sky background was low enough ($\lesssim$12000 counts) to avoid pushing the total raw flux in pixels containing bright (2MASS-detected) stars into the regime where the detector response becomes non-linear. 
\newline

\noindent Both of these nights the weather suffered from very high wind speeds and patchy cirrus cloud cover, with relatively poor seeing ($\sim$0.6-0.9" in the near-IR). The strong winds made re-acquisition of the guide stars challenging after each dither move, resulting in numerous individual exposures that had to be thrown out due to severe jitter from the guide camera attempting to re-acquire through the wind. We reduced these data using standard \texttt{IRAF} \texttt{noao.imred.ccdred} packages and calibrated the final stacked imaging using stars that were detected in 2MASS \citep{skrutskie2006} while also being faint enough to avoid non-linearity (or saturation) in the raw FourStar frames. 
\newline

\begin{figure*}
    \centering
    \includegraphics[width=\linewidth]{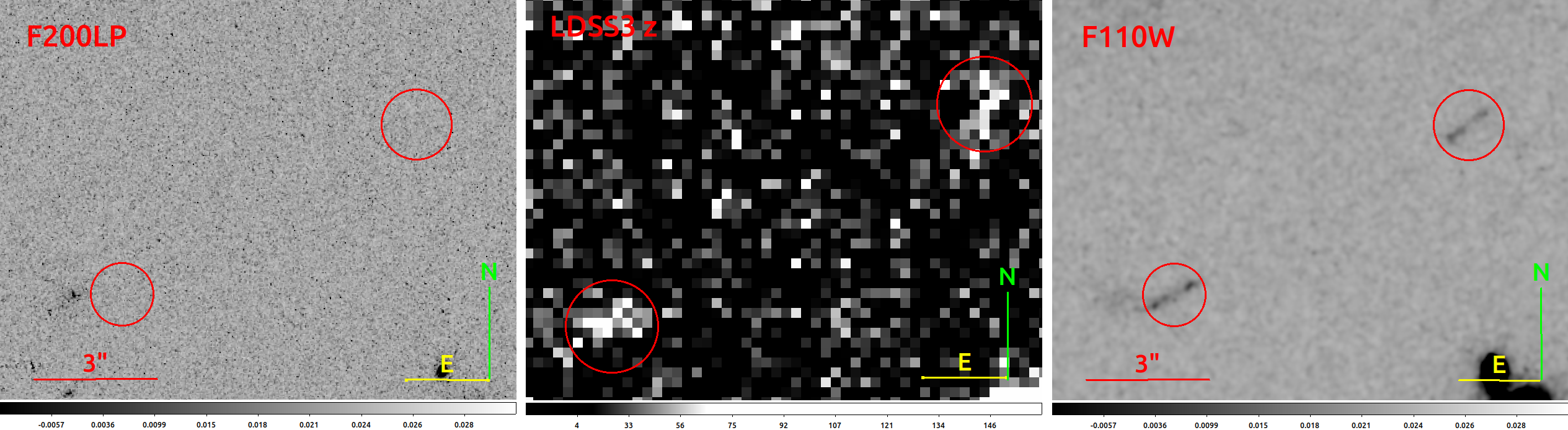}
    \caption{Multiband thumbnail imaging of SPT0216-DO. The source exhibits strong dropout behaviour, with rest-frame emission (indicated by the red circles) being well-detected in the F110W (rightmost) and z-band (center) bandpasses but not in the F200LP (leftmost).}
    \label{fig:0216_cutout}
\end{figure*}

\begin{figure*}
    \centering
    \includegraphics[width=\linewidth]{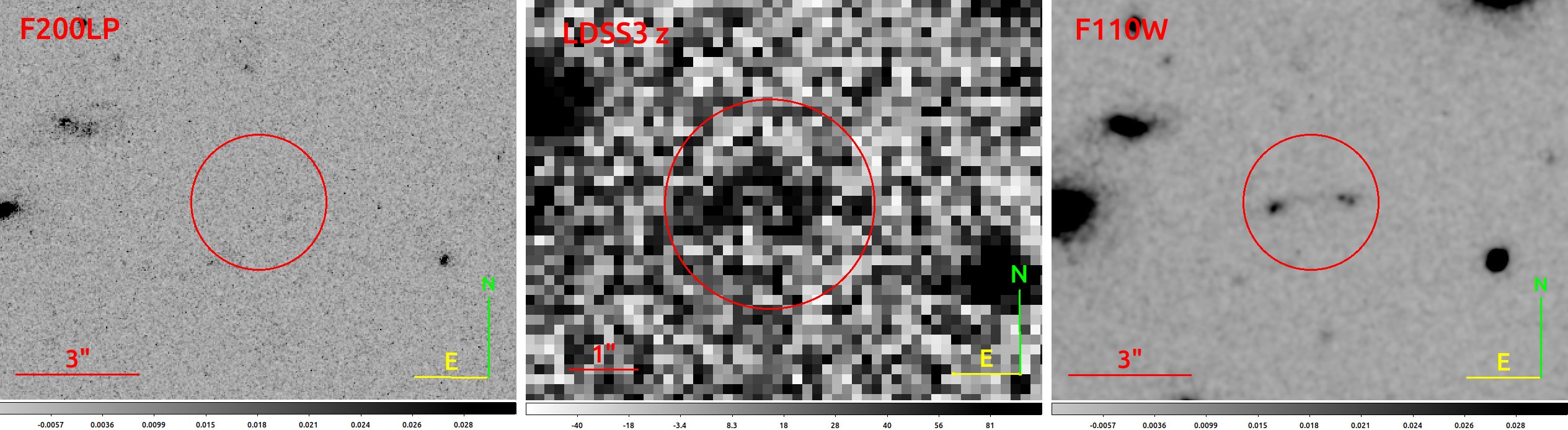}
    \caption{As Figure \ref{fig:0216_cutout} for SPT0304-DO. The source likewise exhibits strong dropout behaviour, being well detected in the F110W (rightmost) and z-band (center) bandpasses, and not detected in the F200LP (leftmost).}
    \label{fig:0304_cutout}
\end{figure*}

\begin{figure*}
    \centering
    \includegraphics[width=\linewidth]{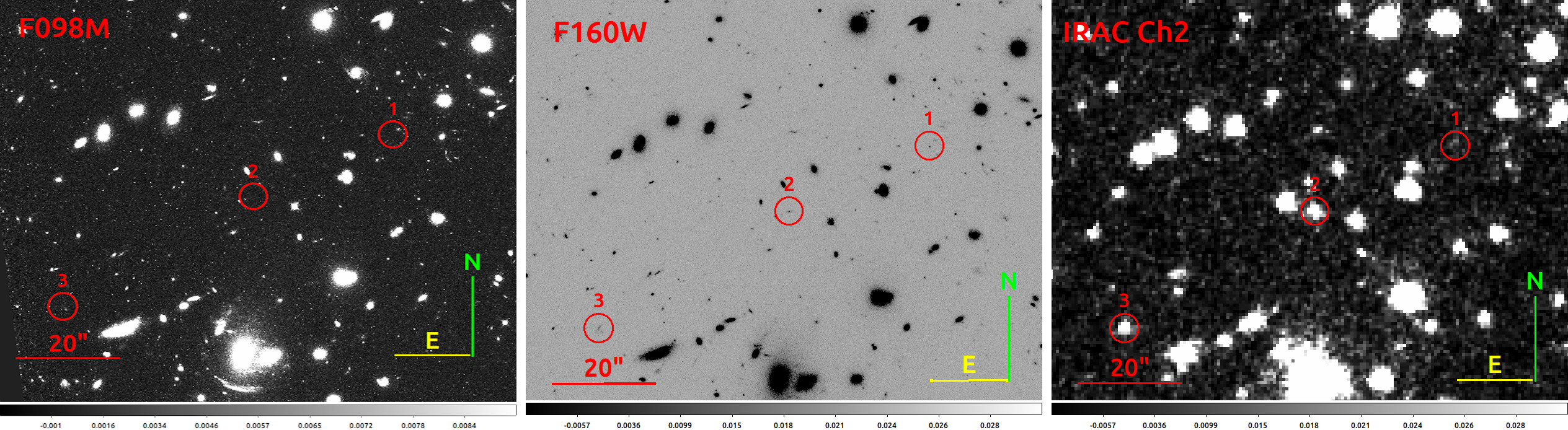}
    \caption{As Figure \ref{fig:0216_cutout} for the three dropouts in the SDSS field. The positions of DO-1, DO-2, and DO-3 are indicated by the label next to each circle.}
    \label{fig:sdss_cutout}
\end{figure*}

\noindent Ultimately, these observations resulted in usable data with total integration times of 2183 s in H and 2696 s in K$_{\rm s}$ for SPT0216-DO, and total integration times of 2017 s in H and 1956 s in K$_{\rm s}$ for SPT0304-DO. The final stacked images have median seeing values of $\sim$0.7 (0.8) in H and $\sim$0.75 (0.8) in K$_{\rm s}$ for SPT0216-DO (SPT0304-DO). Neither dropout candidate is detected in any of these images, so we compute upper limits on their photometry based on the background noise properties of the data.  All photometry was aperture corrected using measurements of bright but unsaturated stars in the respective image fields. The final photometry values across all the observed filter bands are shown in Table \ref{tab:drpoout}, with thumbnail images demonstrating the visual dropout nature of each candidate galaxy presented in Figures \ref{fig:0216_cutout}, \ref{fig:0304_cutout}, and \ref{fig:sdss_cutout}.  The upper limits in H are reported in Table~\ref{tab:drpoout}, but the upper limits in K$_{\rm s}$ are neither reported nor used in the analysis that follows because we later obtained significantly deeper K$_{\rm s}$ imaging from Gemini-South, described in the following section. 

\subsection{Gemini-South/FLAMINGOS-2 K$_{\rm S}$-band Imaging}

\noindent We obtained deep K$_{\rm S}$ imaging of SPT0216-DO and SPT0304-DO with the FLAMINGOS-2 instrument on the Gemini South telescope through program GS-2021B-Q-140 (PI:Bayliss). Each source was observed in Band 1 during good conditions (no cloud cover, 70th \%-ile seeing) with a 13-point dither pattern, with four individual exposures of either 11 s or 13 s (depending on sky background levels) at each dither point, resulting in a total of either 572 s or 676 s of total integration time from each execution of the dither pattern. The dither pattern was executed four times for each source, for total science integration times of 2704 s and 2288 s for SPT0216-DO and SPT0304-DO, respectively. The final seeing in the stacked K$_{\rm S}$-band images was 0.18" for both SPT0216-DO and SPT0304-DO.
\newline

\noindent The data were reduced via the DRAGONS reduction pipeline \citep{Labrie23}. Imaging frames were separated into 21 second science frames, 11 second dark frames, and 11 second flat frames as required by the reduction pipeline. Using these frames we created a master dark and master flat field, as well as a bad pixel mask from observatory-provided data to correct for defects within the detector (i.e. dead pixels or bad rows or columns). Certain imaging frames were also affected by the guide camera arm of the telescope protruding into the field of view; for those exposures, custom pixel masks were created to block out the extent of the field of view where the guide arm was present. The dark and flat fields were then stacked together while ignoring any pixels that fell into either of the bad pixel masks in an offset-to-sky sequence, leaving stacked and aligned science images for the two wavelength bands. 

\begin{table*}
    \centering
    \caption{Dropout Photometry}
    \begin{tabular}{cccccccc}\toprule
         Galaxy&  F098M&  F105W&  F125W&  F140W&  F160W&  IRAC Ch1& IRAC Ch2\\\midrule
         SDSS0952-DO-1&  25.62 $\pm$ 0.17&  25.43 $\pm$ 0.12&  24.08(4)&  23.76(2)&  23.45(3)&  ---& ---\\
         SDSS0952-DO-2&  ---&  ---&  25.06(3)&  24.28(1)&  23.94(2)&  21.00(8)& 20.64(7)\\
         SDSS0952-DO-3&  ---&  26.43$_{-0.42}^{+0.70}$&  25.23(4)&  24.54(2)&  24.23(3)&  ---& 22.60$_{-0.26}^{+0.35}$\\\midrule
 Galaxy& F200LP& LDSS z& F110W& H & Ks& IRAC Ch1&\\ \midrule
 SPT0216-DO&  $>$28.69& 24.82$_{-0.26}^{+0.34}$& 23.87(7)& $>$24.23& $>$24.05& ---&\\
 SPT0304-DO& $>$ 28.78& 24.73$_{-0.25}^{+0.33}$& 23.87$_{-0.09}^{+0.10}$& $>$ 23.9& 23.96$_{-0.28}^{+0.39}$& 23.49$_{-0.36}^{+0.55}$&\\ \bottomrule
    \end{tabular}
    \\[5pt]Photometric $AB_{mag}$ measurements for all observed bandpasses. Upper limits are reported as 1-$\sigma$ (68\% confidence) limits on the flux density.
    \label{tab:drpoout}
\end{table*}

\section{Analysis and Modeling}
\subsection{Lens Modeling}\label{lens model}

\noindent We use pre-existing lens models for the three SDSS galaxies as presented in \citet{Sharon20}. We computed tentative lens models for SPT0216-DO and SPT0304-DO based on the HST-SNAP data described in Section \ref{hstsnap}. We used the public lens modeling software \texttt{lenstool} \citep{Jullo_2007}; this software uses Markov Chain Monte Carlo (MCMC) to explore the parameter space and identify the best fit set of parameters, which results in the smallest scatter between observed and predicted positions of images of the lensed sources. We parameterized each lens with dual pseudo-isothermal ellipsoid (dPIE, \citet{elias07} halos. The parameters of these halos are position $x$, $y$; ellipticity $e$; position angle $\theta$; core radius $r_{core}$; cut radius $r_{cut}$ and normalization $\sigma_0$. We identified cluster-member galaxies based on their red-sequence color \citep{Gladders00}, and assigned a dPIE halo to each, with parameters scaled to its luminosity using scaling relations. We assigned one cluster-scale halo to each lens, with most of its parameters free (the cut radius cannot be constrained by strong lensing, and is fixed at 1500 kpc). We assumed a redshift of the arcs based on the photometric redshift. 
\newline

\noindent Because no other secure multiply imaged objects were observed in the SNAP data, the lens models only used the dropout images themselves as constraints and are thus severely limited. The under-constrained models predict a wide magnification distribution at the location of each arc: for SPT0216-DO, the model predicts a magnification factor $\mu$ between 2.2-275, with 68\% of the results between 5.4 – 22; for SPT0304-DO the model predicts a magnification factor $\mu$ between 10-800 with 68\% of the results between $\sim$ 20-100. We expect that adding imaging data from \textit{JWST} will dramatically improve the accuracy of the lens model, even over short integration times for each target; the combination of \textit{Hubble} and \textit{JWST} imaging has been shown to be effective in improving the geometric and magnification constraints on lens models for other lensing clusters \citep{Mahler23,Diego23,Diego25}.  

\subsection{GALFIT Modeling}
\begin{figure}
    \centering
    \includegraphics[width=\linewidth]{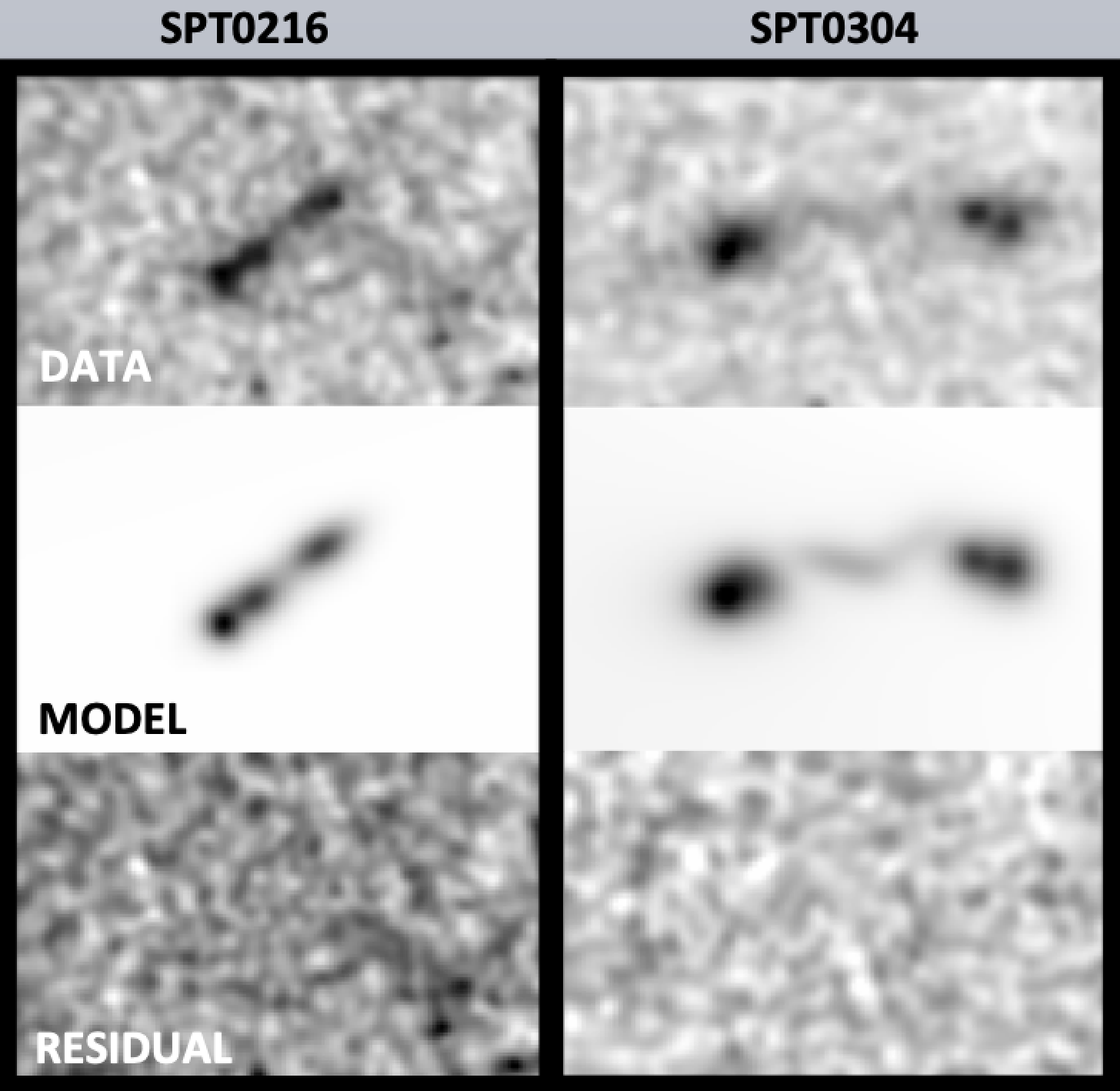}
    \caption{GALFIT models of the SPT targets showing the
spatially resolved clump structures in each dropout galaxy.}
    \label{fig:galfit}
\end{figure}

\noindent Morphological modeling of the lensed arcs was completed using GALFIT \citep{galfit3}. A reference PSF for each modelled band was constructed from an average of moderately-bright stars visible in the F110W image, that were shifted, stacked, and azimuthally filtered to suppress any adjacent objects. Each arc was modelled as a set of Sérsic profiles, with the number of profiles set such that yielded no visible structured residuals at the location of each arc; the modelled arcs and corrected residuals are presented in Figure \ref{fig:galfit}. The image of SPT0216-DO is well fit by three components, each of which is unresolved radially. Each component has an apparent size of no larger than $\sim 0.3$ kpc for each component based on the point-spread function of the F110W filter. Likewise the image of SPT0304-DO is fit by three components that appear to be part of a merging pair. In this case the point-spread function suggests a larger apparent size of $\sim 0.8$ kpc. Neither of these apparent sizes are corrected for lensing magnification correction. The lens models that correspond to these data are currently highly uncertain; accounting for magnification, the sizes measured would be corrected down by a factor of $\sqrt{\mu}$ to true sizes of $18 - 202$ pc for SPT0216-DO and $28-253$ pc for SPT0304-DO (assuming either maximum or minimum magnification). 

\subsection{SED Fitting}\label{fitting}

\noindent A problem inherent to fitting galaxy spectral energy distributions (SEDs) from photometric data is the difficulty in separating bright but more distant galaxies from closer galaxies that are obscured by dust or smaller and thus inherently dimmer. This issue is exacerbated in the case of the dropout candidates due to the upper limits and large error bars on some of the photometric observations. Accurately identifying actual high-$z$ galaxies from the dropout candidates requires fitting synthetic model spectra that can account for a diverse array of stellar populations, star formation histories, and dust attenuation curves within the parameter space. We use the MCMC-based stellar population synthesis code \texttt{Prospector} as the framework for this modeling pipeline \citep{Prospector21}. Models were based off of the MILES stellar population models and MIST isochrones that the Python-FSPS framework provides by default, with nebular continuum and line emission models derived from Cloudy tables \citep{Conroy09,fsps2,Byler17}.
\newline

\begin{table*}
    \centering
    \caption{SED Fitting Priors}
    \begin{tabular}{ccc}\toprule
         Parameter&  Description& Priors\\\midrule
         $z$&  Galactic redshift& Top hat: [1,8]\\
         $t_{age}$&  Galactic age in Myr& Top hat: [0, $t_{univ}$] \\
         A($\lambda$)&  Dust attenuation at wavelength $\lambda$& Top hat: [0,5]\\
         $M/M_{\odot}$&  Total stellar mass formed& Log uniform: [1e9, 5e12]\\
         log($Z/Z_{\odot}$)&  Stellar metallicity& Top hat: [-2, 0.3]\\
         log($U$)&  Gas ionization parameter& Top hat: [-4, -1]\\
 $\tau$& e-folding time for tau decay models&Log uniform: [1e-1, 1e4]\\ \bottomrule
    \end{tabular}
    \label{tab:sps priors}
\end{table*}

\noindent We choose to fit the SED models to the observed photometry, rather than a magnification-scaled version based on the currently available lens models. This choice assumes no correlations or coupled parameters between magnification-dependent parameters, such as stellar mass and star formation rate, and magnification-independent parameters such as dust extinction or redshift. In accordance with the goal of exploring as wide a parameter space as possible, we fit all the models with two star-formation histories: a simple stellar population (SSP) model, wherein the star formation history is a delta function in time and metallicity, and a parameterized exponential tau-decay model, wherein the star formation history is modelled as SFR $\sim e^{-t_{age}/\tau} $. A nebular continuum with simulated nebular line emission is present in all models. However, models were run with and without the Lyman-alpha emission line present: the presence or absence of the Lyman-alpha within each model created a degeneracy in the resultant redshift fits, wherein the Lyman-alpha flux falling into the photometric bandpass would lead to a slightly higher redshift galaxy when compared to a galaxy without the Lyman-alpha flux. 
\newline

\begin{figure*}
    \centering
    \includegraphics[width=0.875\linewidth]{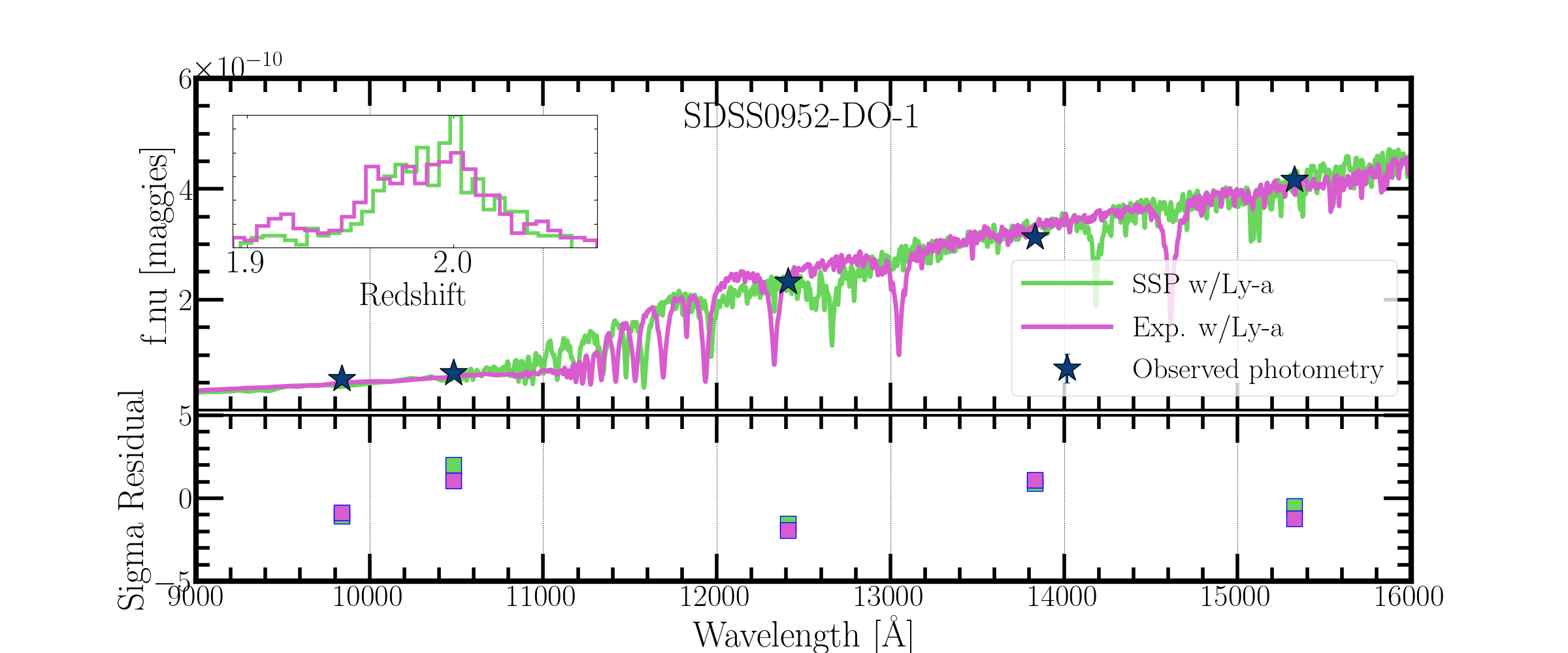}
    \includegraphics[width=0.85\linewidth]{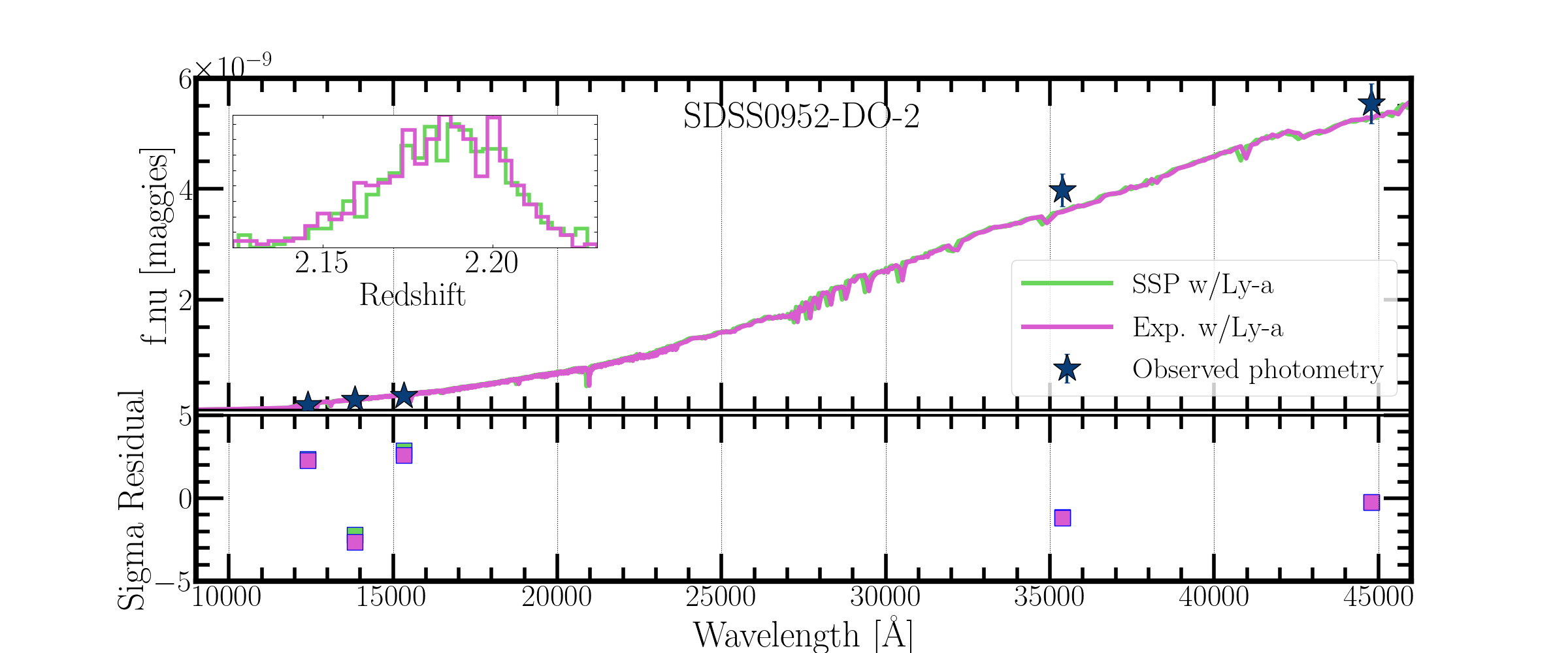}
    \includegraphics[width=0.85\linewidth]{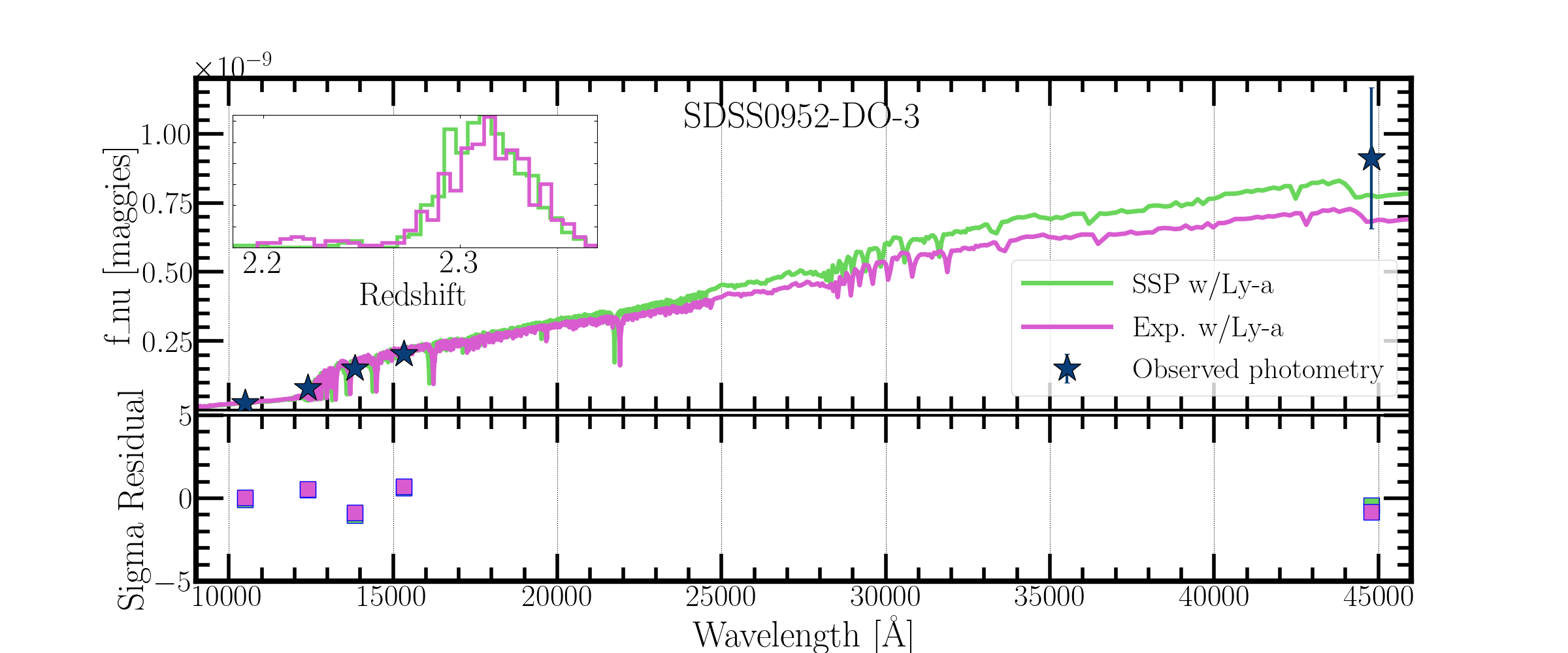}
    \caption{\texttt{Prospector} fits for the SDSS dropout candidates, with the probability density of the fitted redshift plotted as insets. Photometric error bars are included in every plot but are too small to be visible for some sources. The `SSP' and `Exp.' terms within the legend refer to different fitted star-formation histories as detailed in Section \ref{fitting}. Residuals for the photometric data are plotted in green (SSP) and magenta (Exp.). }
    \label{sdss seds}
\end{figure*}

\begin{figure*}
    \centering
    \includegraphics[width=0.85\linewidth]{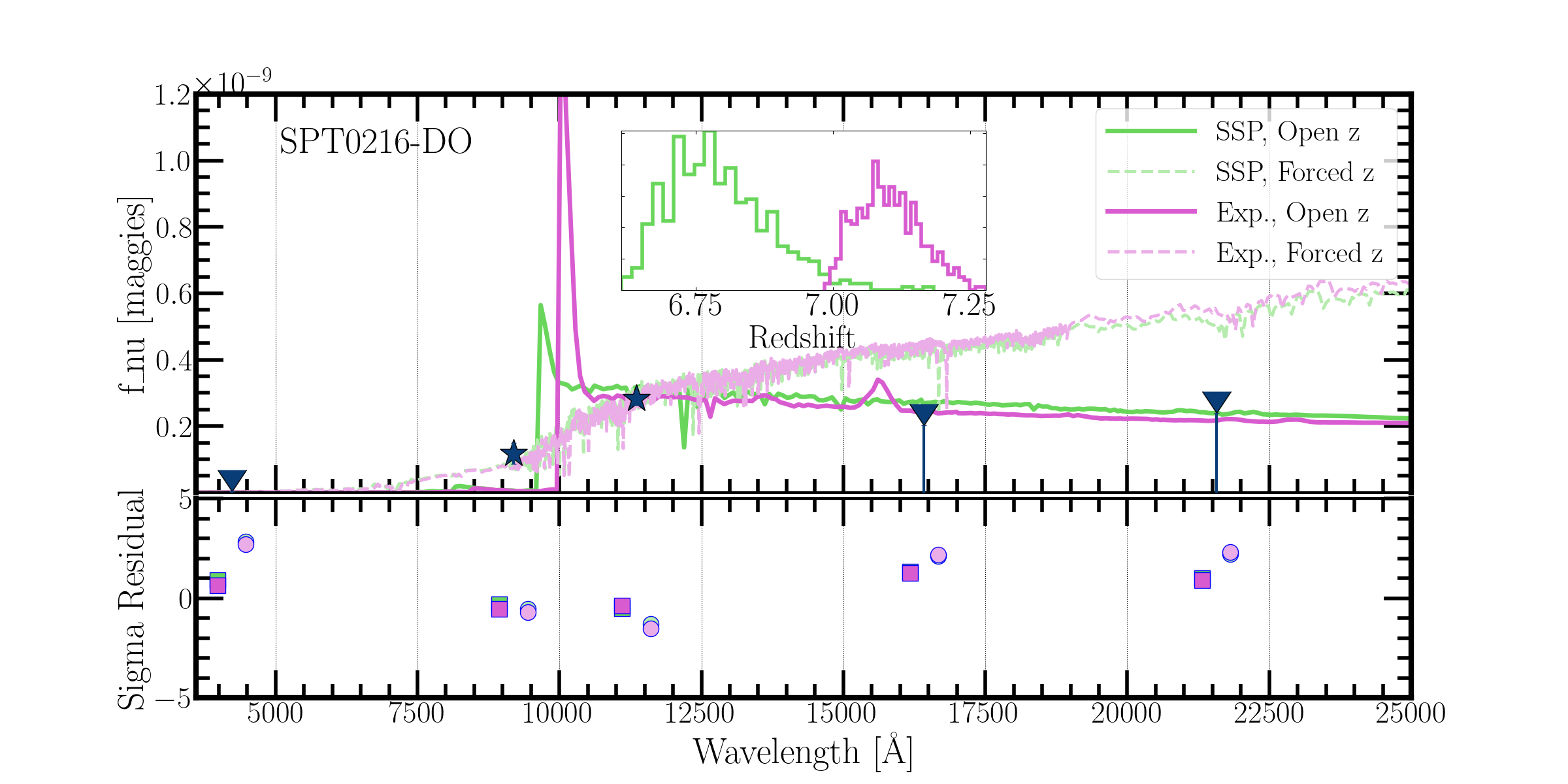}
    \includegraphics[width=0.85\linewidth]{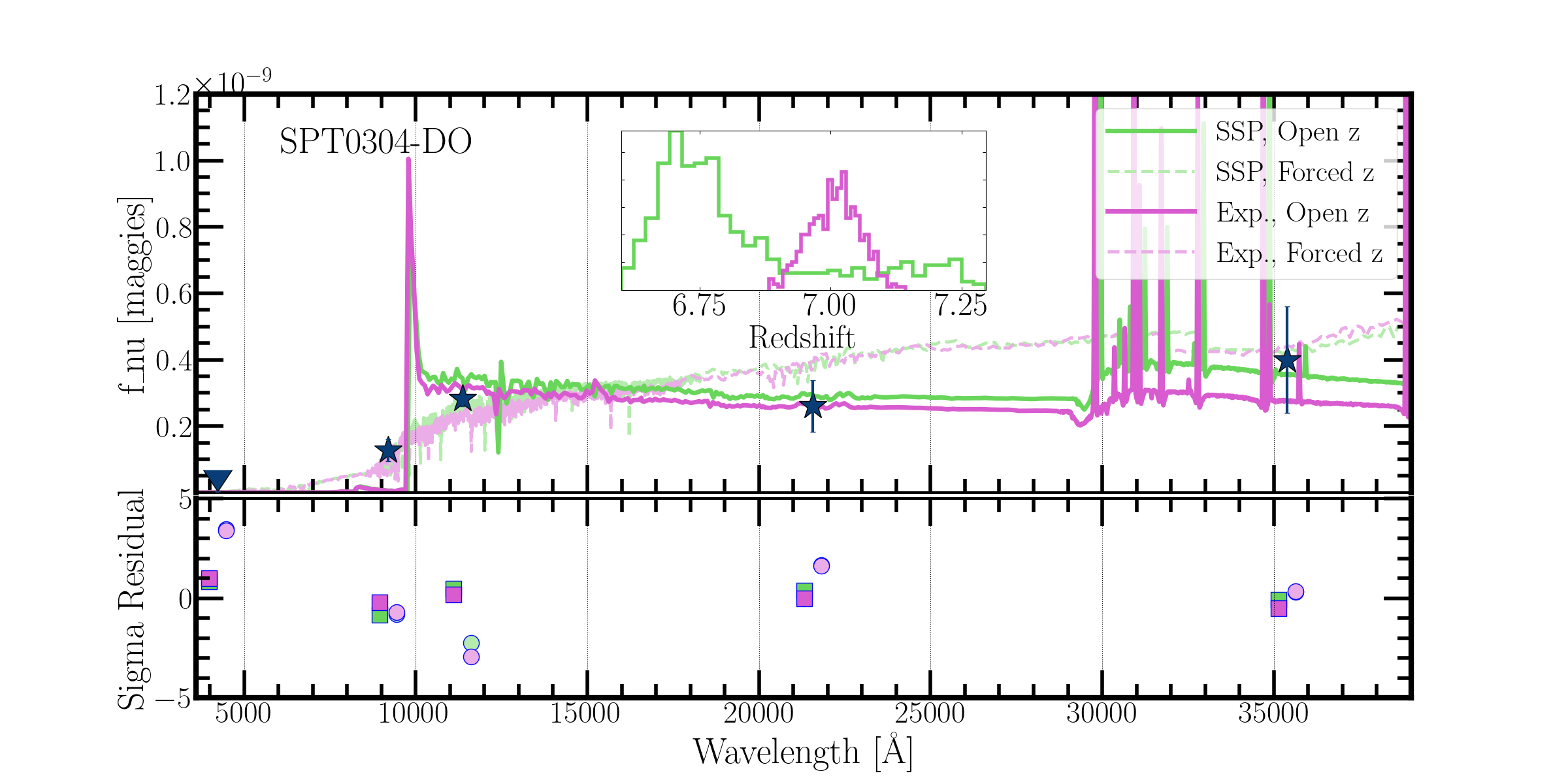}
    \caption{As Figure \ref{sdss seds} for the SPT dropout candidates. Photometric upper limits are displayed as downward facing carats. The solid vs dashed SED plots represent the 'open redshift' vs 'forced redshift' models detailed in Section \ref{fitting} respectively. The photometric residuals for the 'open redshift' (squares) and forced redshift (circles) are offset by 500 \AA ngstrom for the sake of visual clarity. We note that the forced redshift models have higher residuals than the $z \sim 7$ model fits.}
    \label{dropout fits}
\end{figure*}

\noindent All combined this led to a total of four models for each galaxy corresponding to each possible combination of star-formation history (SSP vs exponential tau-decay) and Lyman-alpha (emission vs fully attenuated). Within each of these four models, redshift, stellar metallicity log($Z_{*}/Z_{\odot}$), total stellar mass log($M_{*}/M_{\odot}$), gas ionization parameter \textit{U}, and dust attenuation using the Calzetti et al. attenuation model, were all fit as free parameters.  We make note of the special parameter $t_{univ}$ set as the upper limit of the age prior: to avoid non-physical SED models where a galaxy could be built with stellar populations older than the age of a high-redshift Universe the sampling chain was exploring, we tie the maximum possible age of any search to $t_{univ}$ = the age of the Universe at the corresponding redshift and then parameterize the age as some fraction of that maximum age, ranging from [0,1] inclusive.  The initial set of priors for these parameters are detailed below in Table \ref{tab:sps priors}. 
\newline

\noindent While this model space was extended to search over as wide a range as possible, each separate model necessitated a set of parameter values to serve as an initial prior. In each case, a sample selection of models was created using educated predictions based on the shape of the existing photometric data. These models were compared to the photometry manually, and the set of priors that most closely matched the shape of the data was fed into \texttt{Prospector} as the starting point for the full model to begin its search. The open-source Python module \textit{dynesty} was then used to estimate the Bayesian posteriors using dynamic nested sampling, a method of sampling that utilizes algorithmic methods from both MCMC and nested sampling fitting methods to allow for the creation of complex multi-modal probability distributions \citep{Speagle20}. This final set of posteriors was returned to \texttt{Prospector} and the resultant SED taken as the `true' characteristic spectrum for each galaxy; these SED models are presented in Figures \ref{sdss seds} and \ref{dropout fits}.
\newline

\noindent The primary goal of this work was to select out high-$z$ lensed galaxies as targets for future in-depth spectroscopic study, and with that in mind the initial redshift range was set to explore all possible redshift solutions in order to catch any potential bi-modalities or tails of extreme high- or low-$z$ fits to the sources. After factoring in the initial predictions detailed above, the three SDSS galaxies were identified as highly likely to be $z \sim 2$ galaxies, while the SPT galaxies were identified as highly likely to be $z \sim 7$ galaxies. Once the broad models arrived at robust photo-$z$ solutions, we ran `zoomed in' models in the predicted regions of parameter space in order to more densely sample the parameter ranges and produce final parametric constraints with confidence intervals. 
\newline

\noindent Additionally, two follow-up checks were done to corroborate the parameterization of the derived SED models; firstly, we independently checked the redshift classifications with another stellar population synthesis code, \texttt{Bagpipes} \citep{Carnall18,Carnall19}, searching over an effectively identical parameter space. Because the stellar populations and libraries internal to \texttt{Bagpipes} differ from those used within \texttt{Prospector}, we feel confident that any assumptions inherent to our initial set of models are not strongly biasing the parameter constraints, including the apparent high-$z$/low-$z$ split within the sample of five galaxies. Secondly, we ran a set of Prospector fits forcing the SPT galaxies into a lower redshift space ($z < 5$) while leaving the rest of the parameter space open, as shown in Figure \ref{dropout fits}; in all cases, the resultant SEDs were a noticeably worse fit to the existing data, with higher photometric residuals and higher $\chi^{2}$ values when compared to the high-$z$ fits (SPT0216-DO: $\chi^{2}_{open}$ = 0.431, $\chi^{2}_{forced}$ = 3.593; SPT0304-DO: $\chi^{2}_{open}$ = 0.101, $\chi^{2}_{forced}$ = 0.838). 
\newline

\noindent Based on these results, we conclude that SPT0216-DO and SPT0304-DO are very likely to be true young, unobscured $z \sim 7$ galaxies. The full sets of best-fit values for the successfully fitted parameters are presented in Tables \ref{tab:ssp_fits} and \ref{tab:tau_fits}; we note that no meaningful constraints were able to be obtained for the stellar metallicity or gas ionization parameters. The mass values reported are for a model of the galaxies without taking their magnification into account, and thus all values include some magnification factor $\mu$ as a multiplicative factor to the actual mass value.
\newline

\begin{table*}
    \centering
    \caption{SSP SFH Parametric Fits}
    \begin{tabular}{cccccc}\toprule
         Galaxy&  $z$&  $t_{age}$&  A($\lambda$)& log($\mu M/M_{\odot}$) &$\chi^{2}/DoF$\\\midrule
         SDSS0952-DO-1&  $1.96_{-0.10}^{+0.08}$&  $507_{-162}^{+347}$&  $1.21_{-0.53}^{+0.67}$& $10.95_{-0.15}^{+0.14}$ &0.004\\
         w/o Lyman $\alpha$&  $1.95_{-0.10}^{+0.08}$&  $564_{-234}^{+349}$&  $1.20_{-0.53}^{+0.65}$& $10.94_{-0.14}^{+0.14}$ &0.004\\
         SDSS0952-DO-2&  $2.16_{-0.05}^{+0.06}$&   $151_{-17}^{+23}$&  $3.38_{-0.13}^{+0.31}$& $11.48_{-0.13}^{+0.13}$ &0.004\\
         w/o Lyman $\alpha$&  $2.16_{-0.05}^{+0.06}$&  $151_{-18}^{+25}$&  $3.37_{-0.13}^{+0.31}$& $11.48_{-0.13}^{+0.14}$ &0.002\\
         SDSS0952-DO-3&   $2.28_{-0.10}^{+0.07}$&  $449_{-123}^{+145}$&  $1.06_{-0.50}^{+0.50}$& $10.73_{-0.16}^{+0.15}$ &0.024\\
         w/o Lyman $\alpha$&  $2.28_{-0.10}^{+0.07}$&  $497_{-136}^{+154}$&   $1.06_{-0.51}^{+0.51}$& $10.73_{-0.16}^{+0.15}$ &0.012\\
         SPT0216-DO&  $6.59_{-0.25}^{+0.23}$&  $34_{-18}^{+39}$&  $0.22_{-0.16}^{+0.29}$& $10.74_{-0.55}^{+0.49}$ &1.646\\
 w/o Lyman $\alpha$&  $6.59_{-0.24}^{+0.22}$&  $20_{-10}^{+14}$& $0.22_{-0.16}^{+0.28}$&$10.73_{-0.54}^{+0.49}$ &1.533\\
         SPT0304-DO&  $6.67_{-0.22}^{+0.21}$&  $18_{-14}^{+15}$&  $0.22_{-0.16}^{+0.26}$& $10.37_{-0.40}^{+0.28}$ &0.014\\
         w/o Lyman $\alpha$&   $6.65_{-0.20}^{+0.20}$&  $26_{-11}^{+21}$&  $0.21_{-0.15}^{+0.26}$& $10.40_{-0.33}^{+0.27}$ &0.118\\ \bottomrule
    \end{tabular}
    \\[5pt]Parametric fits for the five dropout candidates using a simple starburst star formation history.
    \label{tab:ssp_fits}
\end{table*}

\begin{table*}
    \caption{Exponential Tau-Decay SFH Parametric Fits}
    \centering
    \begin{tabular}{ccccclc}\toprule
         Galaxy&  $z$&  $t_{age}$&  A($\lambda$)& log($\mu M/M_{\odot}$) &log($\tau$) &$\chi^{2}/DoF$\\\midrule
         SDSS0952-DO-1&  $1.95_{-0.10}^{+0.08}$&   $600_{-237}^{+334}$&  $1.20_{-0.52}^{+0.65}$& $10.94_{-0.15}^{+0.14}$& $0.23_{-0.82}^{+0.84}$ &0.005\\
         w/o Lyman $\alpha$&  $1.94_{-0.08}^{+0.07}$&  $1098_{-280}^{+473}$&  $1.63_{-0.67}^{+0.87}$&  $11.08_{-0.17}^{+0.17}$ &$-0.34_{-0.39}^{+0.91}$ &0.005\\
         SDSS0952-DO-2&  $2.16_{-0.05}^{+0.06}$&   $154_{-18}^{+25}$&  $3.38_{-0.13}^{+0.30}$& $11.48_{-0.13}^{+0.12}$&$0.23_{-0.83}^{+0.83}$ &0.004\\
         w/o Lyman $\alpha$&  $2.19_{-0.05}^{+0.05}$&  $474_{-72}^{+84}$&  $3.01_{-0.20}^{+0.20}$ & $11.53_{-0.08}^{+0.08}$&$0.54_{-0.63}^{+0.40}$ &0.002\\
         SDSS0952-DO-3&   $2.28_{-0.10}^{+0.07}$&  $477_{-126}^{+143}$&  $1.05_{-0.50}^{+0.51}$& $10.73_{-0.16}^{+0.15}$&$0.25_{-0.84}^{+0.83}$ &0.108\\
         w/o Lyman $\alpha$&  $2.27_{-0.09}^{+0.08}$&  $883_{-158}^{+175}$&   $0.97_{-0.45}^{+0.43}$& $10.78_{-0.14}^{+0.14}$&$-0.60_{-0.25}^{+0.25}$ &0.024\\
         SPT0216-DO&  $6.95_{-0.20}^{+0.18}$&  $118_{-76}^{+232}$&  $0.28_{-0.20}^{+0.33}$& $10.58_{-0.47}^{+0.44}$&$0.65_{-0.99}^{+0.92}$ &1.505\\
 w/o Lyman $\alpha$&  $6.67_{-0.20}^{+0.19}$&  $190_{-119}^{+253}$&  $0.27_{-0.19}^{+0.32}$&$10.60_{-0.42}^{+0.41}$&$0.70_{-0.93}^{+0.87}$  &1.654\\
         SPT0304-DO&  $6.98_{-0.17}^{+0.16}$&  $373_{-203}^{+278}$&  $0.20_{-0.14}^{+0.21}$& $10.44_{-0.36}^{+0.26}$ &$1.67_{-1.55}^{+1.57}$ &0.067\\
         w/o Lyman $\alpha$&   $6.67_{-0.19}^{+0.19}$&  $362_{-206}^{+272}$&  $0.21_{-0.14}^{+0.22}$& $10.49_{-0.35}^{+0.25}$&$1.73_{-1.56}^{+1.56}$ &0.337\\ \bottomrule
    \end{tabular}
     \\[5pt]Parametric fits for the five dropout candidates using an exponential tau-decay star formation history.
    \label{tab:tau_fits}
\end{table*}

\section{Summary and Conclusions}

\begin{figure*}[h!]
    \centering
    \includegraphics[width=\linewidth]{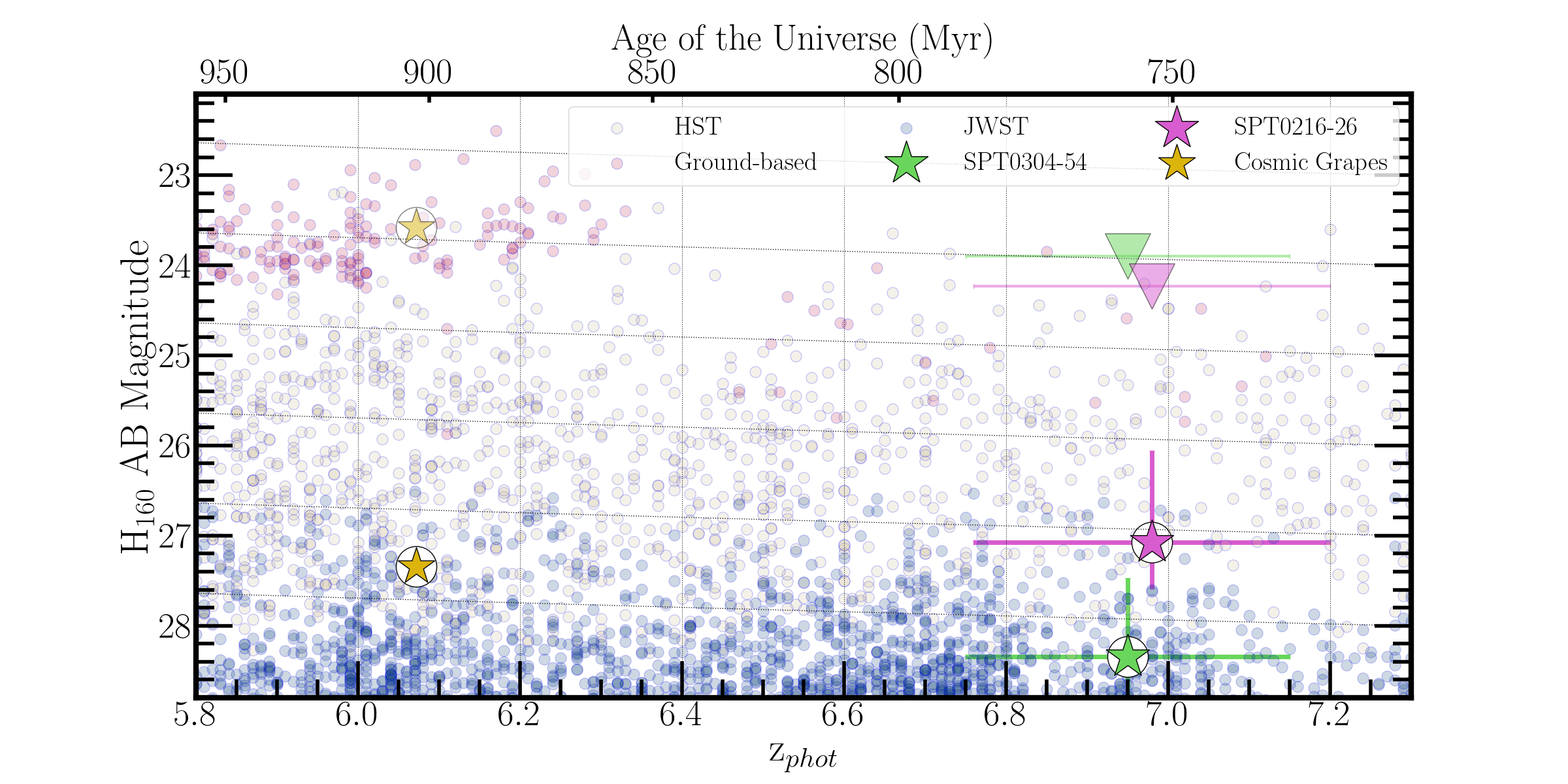}
     \caption{Photometric redshift vs H-band magnitude for high-z galaxy samples from both ground- and space-based surveys. Data compiled from \citet{Bowler12,Bradley14,Finkelstein15,Ishigaki18,Salmon20,Endlsey22,rieke23}. SPT0216-DO and SPT0304-DO appear among the brighter EoR galaxies known, but are intrinsically representative of much fainter galaxies that were far more numerous during the EoR. We also plot the Cosmic Grapes \citep{fujimoto25}, the only other comparable highly magnified galaxy at a similar redshift to the dropouts identified in this work, as a comparison point. All galaxies are plotted with their upper-limit (upper, faded) and estimated demagnified (lower, solid) magnitudes - the magnification makes these galaxies $>$ 3 magnitudes brighter observed compared to their intrinsic luminosities.}
    \label{fig:hiz catalogue}
\end{figure*}

\noindent In this work we present the search and selection process for five candidate high-$z$ dropout galaxies, as well the follow-up observations associated with each galaxy. We discussed the SED fitting process to characterize each of these galaxies based on the existing photometric data, and demonstrated that we can state with a reasonable amount of confidence that two of the five galaxies are photometrically confirmed to be $6.6 < z_{phot} < 7$. In the wider context of the Epoch of Reionization, this $z_{phot}$ measurement for these two SPT dropouts marks them as potential high-value targets for future investigation. With the effects of gravitational lensing, they are among the brightest galaxies discovered at that redshift when compared to other ground-based and JWST surveys. However, correcting for magnification suggests that they are inherently dim ($26 <m_{AB} < 29$) low-mass galaxies of exactly the type thought to be the primary driver of reionization in certain SFG-reionization models, as shown in Figure \ref{fig:hiz catalogue}. Better photometric constraints as well as spectroscopic observation for these dropouts will be able to better constrain new iterations of SED modeling, which will in turn be able to inform the physical characteristics and surrounding IGM conditions of what may be a quintessential galaxy within the EoR. These two bright, highly magnified EoR galaxies are prime targets for future detailed studies with facilities like \textit{JWST}, \textit{ALMA}, etc. In the current literature, there is only one other example of a highly magnified high-$z$ galaxy observed at a comparable redshift and magnitude \citep{fujimoto25}; this work triples the number of available sources for this type of in-depth study. 
\newline

\begin{acknowledgments}
We acknowledge the following facilities and instruments for their contributions to the data that compose this paper: \textit{Hubble Space Telescope}/Wide-Field Camera 3, associated with HST-GO-13003, HST-GO-15307 and HST-GO-16017 (PI: Gladders), and HST-GO-14896 (PI: Bayliss); \textit{Spitzer Space Telescope}/InfraRed Array Camera, associated with Programs $\#$70154 (PI: Gladders) and $\#$80012 (PI: Brodwin); \textit{Magellan}/FourStar Infrared Camera; \textit{Magellan}/Low Dispersion Survey Spectrograph-3; and \textit{Gemini-South}/FLAMINGOS-2, associated with Program GS-2021B-Q-140 (PI:Bayliss). This work was supported by grants associated with programs HST-GO-14896, HST-GO-15307, and HST-GO-16017, all of which were awarded by STScI under NASA contract NAS 5-26555. The South Pole Telescope program is supported by the National Science Foundation (NSF) through awards OPP-1852617 and OPP-2332483. Work at Argonne National Lab is supported by UChicago Argonne LLC, Operator of Argonne National Laboratory (Argonne). Argonne, a U.S. Department of Energy Office of Science Laboratory, is operated under contract $\#$DE-AC0206CH11357. 
\end{acknowledgments}

\noindent 

\software{Aperture Photometry Tool \citep{APT},
          Bagpipes \citep{Carnall18,Carnall19},
          DRAGONS \citep{Labrie23}, 
          DrizzlePac \citep{drizzle},
          dynesty \citep{Speagle20},
          GALFIT \citep{galfit1,galfit3},
          lenstool \citep{Jullo_2007},
          Prospector \citep{Prospector}, 
          SAOImageDS9 \citep{ds9},
          Source Extractor \citep{SEx}
          }

\newpage

\bibliography{references}{}
\bibliographystyle{aasjournal}
\end{document}